\documentclass{article}
\usepackage{icrctc07}

\title{The Absolute, Relative and Multi-Wavelength Calibration of the Pierre Auger Observatory Fluorescence Detectors}
\shorttitle{Auger FD calibration}

\authors{R. Knapik$^{1}$, P. Bauleo, B.R. Becker, J. Brack, R. Caruso, C. Delle Fratte, A. Dorofeev, J. Harton, A. Insolia, J.A.J. Matthews, A. Menshikov, F. Ortolani, P. Petrinca, A. Pichel, S. Riggi, M. Roberts, J. Rodriguez Martino, A.C. Rovero, M. Scuderi, A. Tamashiro, D. Torresi, V. Tuci and L. Wiencke for the  Pierre Auger Collaboration$^{2}$ }

\shortauthors{Knapik and et al.}
\afiliations{$^1$Colorado State University Fort Collins, Colorado, USA\\ $^2$Av. San Martin Norte 304 (5613) Malargue, Prov. de Mendoza, Argentina }
\email{knapik@lamar.colostate.edu}

\abstract{Absolute calibration of the Pierre Auger Observatory fluorescence detectors uses a 375 nm light source at the telescope aperture. This end-to-end technique accounts for the combined effects of all detector components in a single measurement. The relative response has been measured at wavelengths of 320, 337, 355, 380 and 405 nm, defining a spectral response curve which has been normalized to the absolute calibration. Before and after each night of data taking a relative calibration of the phototubes is performed. This relative calibration is used to track both short and long term changes in the detector's response.  A cross check of the calibration in some phototubes is performed using an independent laser technique.  Overall uncertainties, current results and future plans are discussed.}

\begin{document}
\maketitle
%Begin the section.
\section{1. Introduction}

In each of the Pierre Auger Observatory fluorescence detector (FD) buildings there are six identical telescopes that each have a 440 pixel camera. Each pixel of the camera has an individual PMT with an FADC readout that must be converted to a light flux in shower reconstruction.  The combined effects of all the detector components (including optical filter transmittance, mirror reflectivity, PMT gain and quantum efficiency,  etc.) are needed to convert the FADC trace to a number of photons incident on the telescope aperture.  An end-to-end technique has been chosen where all the effects from all detector components are taken into account in a single measurement.

\section{2. The drum light source}

\begin{figure}
\begin{center}
\includegraphics [width=0.42\textwidth]{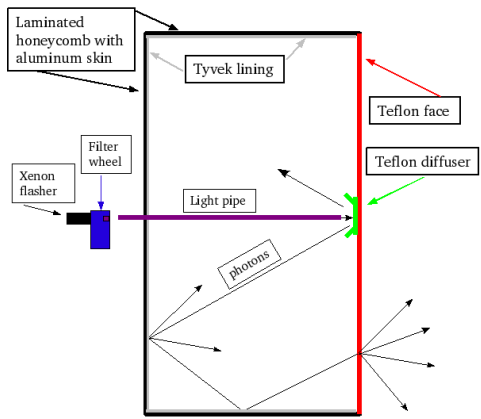}
\end{center}
\caption{The calibration drum.}\label{drum_pic}
\end{figure}

The end-to-end technique involves a portable light source that mounts in the aperture of each FD telescope.  The light source has been designed to uniformly illuminate all 440 pixels in a single camera simultaneously.  The portable light source, referred to as the drum because of its appearance, is a cylinder 1.4 m deep with a 2.5 m diameter shown in figure \ref{drum_pic}.  A light pipe runs from the front face to the back of the drum along the center axis.  The drum can be powered by UV LEDs for the absolute measurement or by a xenon flasher for measurements at different wavelengths.  The broad spectrum of a xenon flasher allows use of notch filters at wavelengths in the 300-420 nm region of interest.  

Where the light pipe meets the front face of the drum a Teflon diffuser directs the light to the walls and back surfaces of the drum which are lined with diffusively reflective materials.  All the photons make multiple bounces to create a surface of uniform illumination in all directions from the face of the drum which is verified with CCD images \cite{astro_ph}.  While perfect drum uniformity is desirable, non–uniformities which are small and well mapped over the surface of the drum are acceptable.  A ray-tracing program using information from the CCD images is used to make corrections for these small non–uniformities.

The absolute calibration of the drum light source is based on a set of Si photodiodes, calibrated at NIST\cite{nist}.  The small surface area and low response of these detectors precludes detection of the small photon flux from the drum surface directly.  A technique has been developed to transfer the low level pulsed drum intensity to a continuous beam three orders of magnitude more intense which is detectable by the calibrated Si photodiode.  The drum is pulsed and the intensity is measured by a reference PMT in a dark room.  On an optical bench the same PMT is placed in front of a variable intensity light source.  This light source is then adjusted to match the recorded drum intensity.  A neutral density filter is then removed, the pulsed light source is changed to a continuous beam and a signal is recorded by the calibrated Si photodiode.  This three step transfer process is described in more detail in reference  \cite{old-icrc-abs}.

\begin{figure}
\begin{center}
\includegraphics [width=0.42\textwidth]{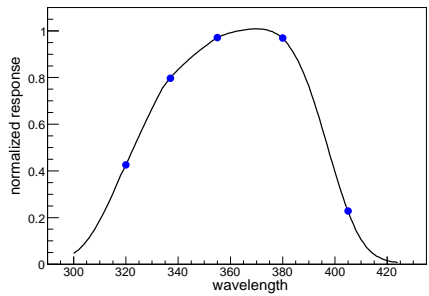}
\end{center}
\caption{The circles are the multi-wavelength measurements normalized to 375 nm.}\label{multi}
\end{figure}

\section{3. Multi-wavelength calibration}

Relative drum intensity measurements at wavelengths of 320, 337, 355, 380 and 405 nm have been made of the drum using the same reference PMT as in the absolute measurements.  The quantum efficiency of the reference PMT has been measured in our lab.  At each wavelength the recorded response from the reference PMT, combined with the quantum efficiency, yield a quantity that is proportional to the number of photons emitted from the drum.  When the drum is placed in the aperture of the FD the signals detected at the various wavelengths combine with the lab work to form a curve of relative camera response shown in figure \ref{multi}.  Interpolation between the measured points is based on a response curve predicted from manufacture specifications.   Corrections made for the filter widths and the statistical uncertainties for the measured points are included in the interpolation.  The relative uncertainty at each wavelength on this curve is 4\%.  This curve is normalized to the absolute measurement made with LEDs at 375 nm.

\begin{figure*}
\begin{center}
\includegraphics [width=1\textwidth]{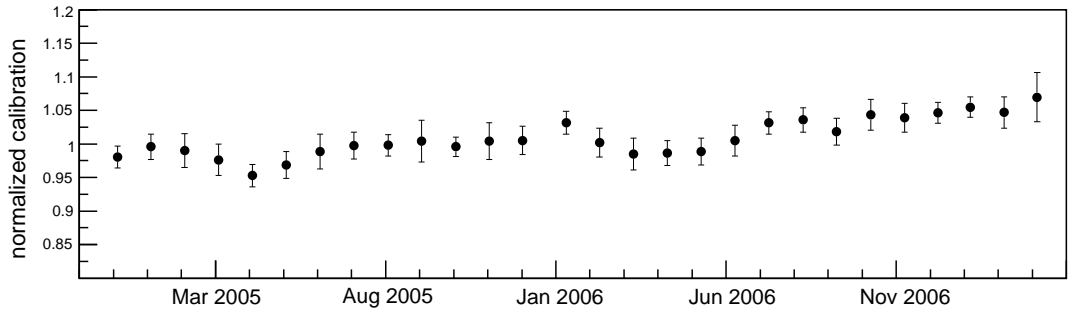}
\end{center}
\caption{The seasonal and long term trends in the response of camera 4 Los Leones.  Each point represents the average camera response for one data taking shift (around 14 nights).  The error bars indicate the one sigma night to night fluctuations.}\label{trends}
\end{figure*}

\section{4. Crosscheck of the absolute calibration}

A portable laser system has been created to provide a cross check of the drum calibration.  The laser is driven out into the field approximately 4 km in front of the FD buildings and fired vertically.  The known Rayleigh scattering cross section of the atmosphere provides a flux of photons which can be predicted when the laser beam energy is well measured by a calibrated energy probe.  This technique includes all the components of the telescope giving an end-to-end calibration with systematic uncertainties that are completely independent from the drum calibration.

Using this method, only a track of pixels can be illuminated at one time.  For other pixels in the camera to be calibrated the laser system has to be moved.  Calibrating all the cameras in this manner is impracticable so only a few pixels in certain cameras are cross checked in this way.  The calibration of the energy probe along with small atmospheric corrections for non perfect Rayleigh scattering are the main sources of uncertainty in this technique.

\begin{figure*}
\begin{center}
\includegraphics [width=1\textwidth]{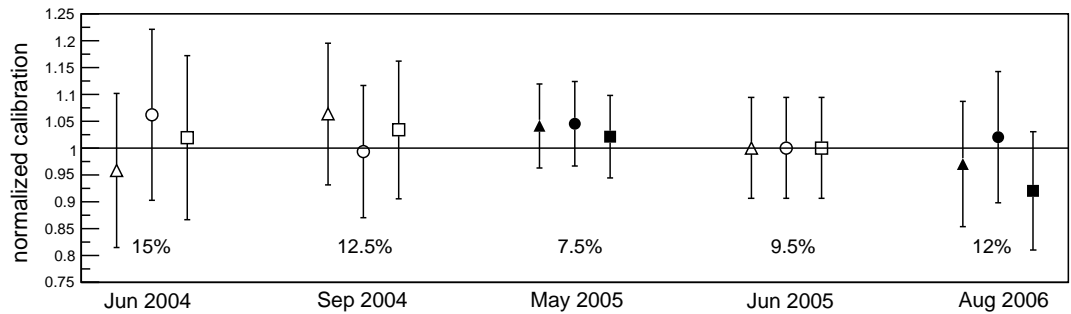}
\end{center}
\caption{
The absolute drum calibrations (hollow shapes) and the laser cross checks (solid shapes) for the past 3 years.  The different shapes represent different cameras. The same three cameras are shown for each of the calibration trips.  All points have been normalized to the June 2005 drum calibration.  The systematic uncertainty for each trip is stated below each group of points.
}\label{abs_plot}
\end{figure*}

\section{5. The relative calibration system}

Before and after each night of data taking a relative calibration is performed.  The system in each FD building uses a set of optical fibers that distribute light from a high powered LED to diffusers located in the centers of all six mirrors that simultaneously illuminate all the pixels\cite{ old_icrc_rel}.

Night to night fluctuations and seasonal trends as shown in figure \ref{trends} have been identified.  The various absolute calibrations and laser cross checks have verified these trends.  In order to reduce systematic uncertainties the relative calibration is used to make corrections to the absolute calibrations.  Figure \ref{abs_plot} shows the consistency of the absolute measurements when the relative calibration is taken into account.  The relative calibration system is used to track any changes in the response of the cameras due to hardware changes and seasonal trends from a reference absolute calibration.

Epochs of time have been defined in the period starting from December 2004 to the present.  In each of these epochs a new reference relative calibration run was chosen to a make an adjustment to the reference absolute calibration.  The epochs span time periods from a few weeks to a few months and have been chosen to correspond to hardware changes.  Any seasonal or long term trends in the response of a camera are taken into account.  There are nightly fluctuations in response that vary at the 3-4\% level in each epoch.  To compare the different drum calibrations and the laser crosschecks the relative calibration information for the specific night in has been used.  Work is being done to fully integrate the relative calibration to provide a correction to the calibration every night of data taking, reducing the impact of night to night fluctuations.

\section{6. Uncertainties and conclusions}

The main sources of uncertainties come from the variable intensity light source used to transfer the drum intensity to the NIST photodiode.  Efforts to reduce the larger sources of uncertainties are ongoing.  Table 1 summarizes the main uncertainties.

\begin{table}[t]
\begin{center}
\begin{tabular}{l|c}
\hline
Drum intensity transfer to  \\ 
calibrated Si photodiode  & 6.0\% \\
NIST calibration\cite{nist}    & 1.5\% \\ 
Temperature effects            & 3.5\% \\ 
Geometrical \\ 
(alignments, areas, etc.)      & 1.8\% \\ 
Reflections \\
(at FD and in lab)             & 1.3\% \\ 
Wavelength distribution effects& 2.5\% \\ 
Drum non-uniformities          & 2.5\% \\ 
Signal readouts \\ 
(currents, FADC traces, etc.)  & 2.3\% \\ 
Camera Response Variations     & 4.0\% \\
\hline
{\bf Total}                    & {\bf 9.5\%} \\ 
\hline
\end{tabular}
 \caption{Table of present uncertianties.}

\end{center}
\end{table}

A redundant lab technique with a different set of systematic uncertainties is in development to provide a cross check to various parts of the current lab procedure.  This setup, along with upgrades in the electronics, will help to reduce the uncertainty on the drum absolute calibration.  In addition, other relative calibration systems will be used to track very long term effects in the mirrors and apertures due to degradation of the optics or possible dust build up.  Full integration of all the relative calibration systems will reduce the number of drum calibrations and laser crosschecks needed.

%This is the reference to .bib file (Without .bib!)
\bibliography{icrc0393}
%This in the bibtex style, is ok.
\bibliographystyle{plain}
\end{document}